\documentstyle[12pt,a4]{article}

\begin{document}

\newcommand{\gf}{G_{\mbox{{\scriptsize F}}}}
\newcommand{\order}{{\cal O}}
\newcommand{\delp}{\Delta P}
\newcommand{\leqsim}{\,\mbox{{\scriptsize $\stackrel{<}{\sim}$}}\,}
\newcommand{\bkk}{B_d\to K^0\bar K^0}
\newcommand{\bksks}{B_d\to K_{\mbox{{\scriptsize S}}}
K_{\mbox{{\scriptsize S}}}}
\newcommand{\bpiphi}{B_s\to\pi^0\Phi}
\newcommand{\bopiphi}{B_s^0\to\pi^0\Phi}
\newcommand{\bbopiphi}{\bar B_s^0\to\pi^0\Phi}
\newcommand{\brhok}{B_s\to\rho^0K_{\mbox{{\scriptsize S}}}}
\newcommand{\bqf}{B_q\to f}
\newcommand{\pcps}{\phi_{\mbox{{\scriptsize CP}}}(B_s)}
\newcommand{\pcpq}{\phi_{\mbox{{\scriptsize CP}}}(B_q)}
\newcommand{\pw}{\phi_{\mbox{{\scriptsize W}}}}
\newcommand{\acp}{a_{\mbox{{\scriptsize CP}}}}
\newcommand{\acpdir}{{\cal A}_{\mbox{{\scriptsize CP}}}^
{\mbox{{\scriptsize dir}}}}
\newcommand{\acpmi}{{\cal A}_{\mbox{{\scriptsize
CP}}}^{\mbox{{\scriptsize mix-ind}}}}
\newcommand{\acc}{A_{\mbox{{\scriptsize CC}}}}
\newcommand{\aew}{A_{\mbox{{\scriptsize EWP}}}}
\newcommand{\heff}{{\cal H}_{\mbox{{\scriptsize eff}}}(\Delta B=-1)}
\newcommand{\xif}{\xi_f^{(q)}}
\newcommand{\xikk}{\xi_{K^0\bar K^0}^{(d)}}
\newcommand{\VmA}{\mbox{{\scriptsize V--A}}}
\newcommand{\VpA}{\mbox{{\scriptsize V+A}}}
\newcommand{\VpmA}{\mbox{{\scriptsize V$\pm$A}}}
\newcommand{\beq}{\begin{equation}}
\newcommand{\eeq}{\end{equation}}
\newcommand{\bea}{\begin{eqnarray}}
\newcommand{\eea}{\end{eqnarray}}
\newcommand{\non}{\nonumber}
\newcommand{\lab}{\label}
\newcommand{\la}{\langle}
\newcommand{\ra}{\rangle}
\newcommand{\np}{Nucl.\ Phys.}
\newcommand{\pl}{Phys.\ Lett.}
\newcommand{\prl}{Phys.\ Rev.\ Lett.}
\newcommand{\pr}{Phys.\ Rev.}
\newcommand{\zp}{Z.\ Phys.}

\setcounter{page}{0}
\thispagestyle{empty}
\begin{flushright}
TUM-T31-76/94\\
September 1994
\end{flushright}
\vspace*{0.2cm}
\begin{center}
\vspace{1.5cm}
{\Large{\bf Mixing-induced CP Violation in the Decay}}\\
\vspace{0.5cm}
{\Large{\bf $\bkk$ within the Standard Model
\footnote[1]{Supported by the German
Bundesministerium f\"ur Forschung und Technologie under contract
06 TM 732.}}}\\
\vspace{2cm}
{\large{\sc Robert Fleischer}}\\
\vspace{0.5cm}
{\sl Technische Universit\"at M\"unchen\\
Physik Department\\
Institut f\"ur Theoretische Physik\\
D-85748 Garching\\
Germany}\\
\vspace{1.5cm}
{\large{\bf Abstract}}\\
\vspace{1cm}
\end{center}

Recently, flavour $SU(3)$ symmetry of strong
interactions has been combined with certain dynamical assumptions to
derive triangle relations among $B$-meson decay-amplitudes. We show
that these relations allow a prediction of the mixing-induced CP
asymmetry $\acpmi(B_d\to K^0\bar K^0)$. Contrary to statements made in
several previous papers, this asymmetry should be non-vanishing in the
Standard Model due to QCD-penguins with internal up- and charm-quark
exchanges and could be as large as $\order(30\%)$. The branching ratio
BR$(B_d\to K^0\bar K^0)$ is expected to be of $\order(10^{-6})$.
In the future, the results presented in this letter should allow
interesting tests of the $SU(3)$ triangle relations and of the
Standard Model description of CP violation.

\newpage

CP-violating asymmetries in neutral $B$-meson decays are of special
interest for an experimental test of the Cabibbo-Kobayashi-Maskawa--model
(CKM--model) of CP violation \cite{km}. In contrast to
the situation arising in the charged $B$-meson system, where only
{\it direct} CP violation is present, it is a characteristic feature
of the neutral $B$-meson system that also {\it mixing-induced} CP
violation, which is generated by the interference between $B^0-\bar B^0$
mixing and decay processes, may contribute significantly to the
CP-violating asymmetries \cite{dr}-\cite{qui}.

The major point of this letter is a prediction of the mixing-induced
CP asymmetry of the decay $\bkk$ originating from the generic
QCD-penguin process $b\to d\bar ss$. The main inputs are the $SU(3)$
flavour symmetry of strong interactions and certain dynamical
assumptions to be specified below. In the previous literature, it has
been claimed by several authors (see, e.g., refs.~\cite{lp,nir,qui})
that decays such as $\bksks$ or $\bkk$ (the CP asymmetries of both
channels are equal) were useful modes to test the
Standard Model, since it would predict {\it zero} CP-violating
asymmetries due to a cancellation of weak decay- and mixing-phases. We
point out that this statement is only correct, if the QCD-penguin
amplitudes are dominated by internal top-quark exchanges. As we shall
see, however, QCD-penguins with internal up- and charm-quarks may also
play an important role and could lead to rather large CP asymmetries of
$\order(10-50)\%$. Therefore, non-vanishing
CP-violating asymmetries measured in $\bksks$ (or $\bkk$) would not
necessarily give hints to contributions from physics beyond the
Standard Model as emphasized, e.g., in ref.~\cite{qui}. The importance of
pure penguin-induced $b\to d\bar ss$ modes in respect of direct CP violation
has been emphasized previously by G\'erard and Hou \cite{gh}.

If we consider a neutral $B_q$-decay $(q\in\{d,s\})$ into a CP
self-conjugate final state $|f\rangle$, e.g., the transition $\bkk$,
the time-dependent and time-integrated CP asymmetries are given by
\bea
\lefteqn{\acp(t)\equiv\frac{\Gamma(B_q^0(t)\to f)-\Gamma(\bar
B_q^0(t)\to f)}{\Gamma(B_q^0(t)\to f)+\Gamma(\bar
B_q^0(t)\to f)}=}\nonumber\\
&&\acpdir(\bqf)\cos(\Delta M_q
t)+\acpmi(\bqf)\sin(\Delta M_q t)\lab{e1}
\eea
and
\bea
\lefteqn{\acp\equiv\frac{\int_0^\infty dt\left[
\Gamma(B_q^0(t)\to f)-\Gamma(\bar B_q^0(t)\to f)\right]}
{\int_0^\infty dt \left[\Gamma(B_q^0(t)\to f)+\Gamma(\bar
B_q^0(t)\to f)\right]}=}\nonumber\\
&&\frac{1}{1+x_q^2}\left[\acpdir(\bqf)+x_q\acpmi(\bqf)\right]\lab{e2},
\eea
respectively. Here, $\Delta M_q>0$ is the mass splitting of the
physical $B^0_q-\bar B^0_q$ mixing-eigenstates and
$x_q\equiv\tau_{B_q}\Delta M_q$ denotes the so-called mixing-parameter. In
eqs.~(\ref{e1}) and (\ref{e2}), we have separated the {\it direct}
CP-violating contributions characterized by
\beq\lab{e3}
\acpdir(\bqf)\equiv\frac{1-\vert\xif\vert^2}
{1+\vert\xif\vert^2}
\eeq
from those describing {\it mixing-induced} CP violation which are
proportional to
\beq\lab{e4}
\acpmi(\bqf)\equiv\frac{2\mbox{Im}\xif}{1+\vert\xif\vert^2}.
\eeq
The phase convention
independent quantity $\xif$ contains essentially all the
information needed to evaluate the CP-violating asymmetries. It is given by
\beq\lab{e5}
\xif=\exp\left[-i\Theta_{M_{12}}^{(q)}\right]
\frac{A(\bar B_q^0\to f)}{A(B_q^0\to f)},
\eeq
where $A(\bar B_q^0\to f)$ and $A(B_q^0\to f)$ are decay amplitudes and
\beq\lab{e6}
\Theta_{M_{12}}^{(q)}=\pi+2\mbox{arg}(V_{tq}^{\ast}V_{tb})-\pcpq
\eeq
is the $B^0_q-\bar B^0_q$ mixing phase which is a function of the
complex phases of the CKM matrix \cite{km}.
The phase $\pcpq$ arises from our freedom of choosing a CP
phase-convention and is defined by the relation
$({\cal CP})|B_q^0\ra=\exp[i\pcpq]|\bar B_q^0\ra$. In the convention
independent expression (\ref{e5}), $\pcpq$ is cancelled by the ratio of
decay amplitudes.

If a neutral $B_q$-meson decay into a final CP eigenstate is dominated
by a single CKM-amplitude, $\acpdir$ vanishes and the uncertainties
related to unknown hadronic matrix elements cancel in the mixing-induced
asymmetry $\acpmi$. In this very interesting case, $\acpmi$ is
a theoretical clean measure of the angles appearing in the so-called
unitarity triangle (for a recent phenomenological analysis of this
triangle see, e.g.,  ref.~\cite{blo}). An example of such a decay is
the channel $B_d\to\psi K_{\mbox{{\scriptsize S}}}$ which should allow a
very clean determination of the angle $\beta$ in the unitarity
triangle from the CP asymmetry $\acpmi(B_d\to\psi K_{\mbox{{\scriptsize
S}}})=-\sin2\beta$.

On the other hand, if several amplitudes with both different CP-violating
weak and CP-conserving strong phases contribute to a neutral
$B_q$-decay, the hadronic uncertainties do not cancel and
a theoretical clean prediction of the CP asymmetries
(\ref{e1}) and (\ref{e2}) is {\it a priori} not possible. Furthermore, we
expect significant direct CP violation. A decay in this category
is, e.g., the pure penguin-induced mode $\bkk$ \cite{rf1}.
Here, the amplitudes with different weak and strong phases mentioned
above arise from QCD-penguins with internal up-, charm- and top-quark
exchanges.

After this short introduction, let us now come to our main point.
Using $SU(3)$ flavour symmetry of strong interactions
\cite{zep}-\cite{hmt} and certain plausible dynamical assumptions
(e.g., neglect of annihilation topologies), several triangle relations
among $B$-meson decay amplitudes into $\pi\pi$, $\pi K$ and $K\bar K$
final states have been derived in a recent series of interesting
publications \cite{grl}-\cite{lon}. In this letter, our main intention
is to point out that these relations in combination with the recent results
of ref.~\cite{bf} allow an interesting prediction of the
mixing-induced CP-violating asymmetry $\acpmi(\bkk)$.

To this end, let us use the same notation as in
refs.~\cite{grl}-\cite{bf} and
denote the amplitudes corresponding to $b\to d$ ($\bar b\to \bar d$)
and $b\to s$ ($\bar b\to \bar s$)
QCD-penguins generically by $\bar P$ $(P)$ and $\bar P'$ $(P')$,
respectively. Then,
taking into account that $({\cal CP})|K^0\bar K^0\ra=+|K^0\bar K^0\ra$
and applying the Wolfenstein parametrization \cite{wolf} of the
CKM-matrix, we obtain
\beq\lab{e7}
\xikk=-\exp(-i2\beta)\frac{\bar P}{P}.
\eeq
In refs.~\cite{grl}-\cite{lon}, it has been
assumed that the QCD-penguin amplitudes are dominated by internal
top-quark exchanges. However, as has been pointed out in \cite{bf},
sizable contributions may also arise from QCD-penguins with internal
up- and charm-quarks. Including these additional penguin amplitudes,
which will turn out to be essential for the CP-violating effects
arising in the mode $\bkk$, we find \cite{bf}
\beq\lab{e8}
\frac{\bar
P}{P}=\frac{\bar\rho_P}{\rho_P}\exp\left[i(2\beta+\psi-\bar\psi)\right],
\eeq
where
\beq\lab{e9}
\rho_{P}=\frac{1}{R_t}\sqrt{R_t^2-2R_t|\delp|\cos(\beta+\delta_{\delp})+
|\delp|^2}
\eeq
and
\beq\lab{e10}
\tan\psi=\frac{|\delp|\sin(\beta+\delta_{\delp})}
{R_t-|\delp|\cos(\beta+\delta_{\delp})}.
\eeq
The CP-conjugate quantities $\bar\rho_P$ and $\bar\psi$ can be
obtained easily from (\ref{e9}) and (\ref{e10}) by substituting
$\beta\to-\beta$. In eqs.~(\ref{e9}) and (\ref{e10}),
$\delp$ describes the contributions of the
QCD-penguins with internal $u$- and $c$-quarks and is defined by
the ratio of strong penguin-amplitudes \cite{bf}
\beq\lab{e11}
\delp\equiv|\delp|\exp(i\delta_{\delp})\equiv\frac{P_c-P_u}{P_t-P_u},
\eeq
whereas the CKM-factor
\beq\lab{e12}
R_t\equiv\frac{1}{\lambda}\frac{|V_{td}|}{|V_{cb}|}
\eeq
represents one side of the unitarity triangle. Present experimental
data imply that $R_t$ is of $\order(1)$ \cite{blo}. Note that $\delp$ is
affected strongly by hadronic uncertainties, in particular by unknown
strong final state interaction phases. In the limit of degenerate $u$-
and $c$-quark masses, $\delp$ would vanish due to the GIM mechanism.
However, since $m_u\approx4.5$ MeV, whereas $m_c\approx1.3$ GeV, this
GIM cancellation is incomplete and in principle sizable effects
arising from $\delp$ could be expected~\cite{bf}.

Combining (\ref{e7}) and (\ref{e8}) gives the expression
\beq\lab{e1a}
\xikk=-\frac{\bar\rho_P}{\rho_P}\exp\left[i(\psi-\bar\psi)\right].
\eeq
If we consider only QCD-penguins with internal top-quark exchanges
corresponding to $\delp=0$ \cite{grl}-\cite{lon}, the weak decay- and
mixing-phases cancel each other and we get
\beq\lab{e1b}
\xikk(\delp=0)=-1.
\eeq
Consequently, the CP asymmetries (\ref{e1}) and (\ref{e2}) vanish in
this approximation. In several previous papers (see, e.g.,
ref.~\cite{qui}), it has been claimed that this result would provide a
test of the Standard Model and that observed non-vanishing CP
asymmetries would indicate physics beyond the Standard Model. However,
within the Standard Model, non-vanishing asymmetries $\acpdir(\bkk)$
and $\acpmi(\bkk)$ may arise from QCD-penguin contributions with
internal $u$- and $c$-quarks.

In order to illustrate this statement quantitatively, let us apply,
as in ref.~\cite{bf}, the perturbative approach of Bander,
Silverman and Soni~\cite{bss} -- the so-called ``BSS-mechanism'' -- to
estimate $\delp$. We will also investigate the expected
magnitude of the ``average'' branching ratio BR$(\bkk)$ defined by
\bea
\mbox{BR}(\bkk)&\equiv&\frac{1}{2}\int_0^\infty dt
\left[\Gamma(B^0_d(t)\to K^0\bar K^0)+\Gamma(\bar B^0_d(t)\to K^0\bar
K^0)\right]=\nonumber\\
&&\frac{1}{2}\left(1+|\xikk|^2\right)\Gamma(B^0_d\to K^0\bar
K^0)\tau_{B_d}.\lab{e12a}
\eea
Here, $\Gamma(B^0_d\to K^0\bar K^0)$ denotes the transition rate of
the decay $B^0_d\to K^0\bar K^0$ which can be obtained by performing
the usual phase-space integrations.

To simplify the discussion, we neglect the
renormalization group evolution from $\mu=\order(M_W)$ down to
$\mu=\order(m_b)$ and take into account QCD renormalization effects
only approximately through the substitution $\alpha_s\to\alpha_s(\mu)$
(for a discussion of QCD-corrections affecting $\bkk$, see
ref.~\cite{rf1}). Then, choosing $R_t=1$ and various
angles $\beta$, we find the curves shown in
Figs.~1--3 describing the dependences of $\acpmi(\bkk)$,
$\acpdir(\bkk)$ and of the corresponding time-integrated asymmetry
$\acp$, respectively, on the momentum transfer $k^2$ of the gluons
appearing in the usual QCD-penguin diagrams.
Applying, in addition to the BSS-mechanism \cite{bss}, the {\it
factorization} approximation and the form factors presented by Bauer,
Stech and Wirbel \cite{bsw} to estimate the relevant hadronic matrix
elements, we obtain the curves for BR$(\bkk)$ depicted in Fig.~4.
The details of the calculation of $\delp$ within the
perturbative BSS-mechanism can be found in ref.~\cite{bf}, whereas the
evaluation of the branching ratio BR$(\bkk)$ has been outlined in
ref.~\cite{rf1}. In drawing Figs.~1--4, we have taken into
account that the present range of $|V_{ub}/V_{cb}|$ implies
$\beta\leqsim 45^\circ$ \cite{blo}. Looking at these figures and
choosing $k^2$ to lie within the ``physical'' range
\beq\lab{e22}
\frac{1}{4}\leqsim\frac{k^2}{m_b^2}\leqsim\frac{1}{2},
\eeq
which follows from simple kinematical considerations at the quark
level, we expect rather large asymmetries of the order $(10-50)\%$,
which are quite promising from the experimental point of view, and
BR$(\bkk)=\order(10^{-6})$. We are aware of the fact that the
numerical estimates given here are very rough. They illustrate, however,
the expected orders of magnitude.

After this short quantitative illustration, let us now turn to the
prediction of the mixing-induced CP asymmetry $\acpmi(\bkk)$.
Using two rates that determine $|P|$ and $|P'|$, e.g., those of the
modes $B^+\to K^+\bar K^0$ and $B^+\to\pi^+ K^0$, respectively, and
the triangle relations~\cite{hlgr}-\cite{lon}
\beq\lab{e13}
\begin{array}{rcl}
A(B^0_d\to\pi^+\pi^-)+\sqrt{2}A(B^0_d\to\pi^0\pi^0)&=&\sqrt{2}
A(B^+\to\pi^+\pi^0)\\
(T+P)\qquad+\qquad(C-P)&=&(T+C)
\end{array}
\eeq
and
\beq\lab{e14}
\begin{array}{rcl}
A(B^0_d\to\pi^- K^+)/r_u+
\sqrt{2}A(B^0_d\to\pi^0 K^0)/r_u&=&\sqrt{2}
A(B^+\to\pi^+\pi^0)\\
(T+P'/r_u)\qquad+\qquad(C-P'/r_u)
&=&(T+C),
\end{array}
\eeq
where $r_u=V_{us}/V_{ud}$ and $T$ ($C$) refers to the ``tree''
(``colour-suppressed'') amplitude of the decay $B^+\to\pi^+\pi^0$, the
relative angle $\vartheta$ between $P$ and $P'$ can be measured. In
contrast to the assertions made in \cite{hlgr}-\cite{lon}, $\vartheta$
is not equal to the CKM-angle $\beta$, but receives some hadronic
corrections \cite{bf}:
\beq\lab{e15}
\vartheta=\beta+\psi-\psi'.
\eeq
Here, $\psi'$ is a pure strong phase that is given by \cite{bf}
\beq\lab{e16}
\tan\psi'=\frac{|\delp|\sin\delta_{\delp}}
{1-|\delp|\cos\delta_{\delp}}.
\eeq
If we consider in addition the corresponding CP-conjugate processes,
the angle
\beq\lab{e17}
\bar\vartheta=-\beta+\bar\psi-\psi'
\eeq
can be determined as well. Note that $\bar\psi'=\psi'$, since no weak
phases are present in (\ref{e16}). Consequently, combining (\ref{e15}) and
(\ref{e17}) appropriately, we find
\beq\lab{e18}
2\beta+\bar\vartheta-\vartheta=\bar\psi-\psi
\eeq
and $\xikk$ can be expressed as
\beq\lab{e19}
\xikk=-\frac{|\bar P|}{|P|}\exp\left[-i(2\beta+\bar\vartheta-
\vartheta)\right]
\eeq
which yields
\beq\lab{e20}
\acpdir(\bkk)=\frac{|P|^2-|\bar P|^2}{|P|^2+|\bar P|^2}
\eeq
and
\beq\lab{e21}
\acpmi(\bkk)=\frac{2|P||\bar P|\sin(2\beta+\bar\vartheta-\vartheta)}
{|P|^2+|\bar P|^2}.
\eeq
Note that eq.~(\ref{e20}) implies that $\acpdir(\bkk)$ should be equal
to the CP-violating asymmetry
\beq\lab{e20a}
\acp(B^\pm\to K^\pm K^0)\equiv\frac{\Gamma(B^+\to K^+\bar K^0)
-\Gamma(B^-\to K^- K^0)}{\Gamma(B^+\to K^+\bar K^0)
+\Gamma(B^-\to K^- K^0)}.
\eeq
The expression (\ref{e21}) for the mixing-induced CP asymmetry
$\acpmi(\bkk)$ is interesting in two respects:
\begin{itemize}

\item[i)] If $2\beta$ is known, the triangle relations (\ref{e13})
and (\ref{e14}) (and those of the corresponding CP-conjugate processes)
allow a prediction of the CP-violating asymmetry $\acpmi(\bkk)$. In
principle, the angle $\beta$ can be determined from the $SU(3)$
triangle relations as well \cite{hlgr}-\cite{lon}. However,
irrespectively of $SU(3)$-breaking effects and certain neglected
diagrams (annihilation topologies, etc.),
the QCD-penguin contributions with internal $u$- and
$c$-quarks preclude a clean determination of $\beta$ by using the
branching ratios only (see eqs.~(\ref{e15}) and (\ref{e17})).
As has been pointed out in ref.~\cite{bf},
this difficulty can be overcome by measuring in addition the ratio
$x_d/x_s$ of $B^0_d-\bar B^0_d$ to $B^0_s-\bar B^0_s$ mixings
to obtain the CKM-parameter $R_t$. The
probably cleanest way of determining $2\beta$ is the measurement of
the mixing-induced CP asymmetry $\acpmi(B_d\to\psi
K_{\mbox{{\scriptsize S}}})$. An interesting possibility to extract
$\sin2\beta$ from the two branching ratios BR$(K^+\to\pi^+\nu\bar\nu)$
and BR$(K_{\mbox{{\scriptsize L}}}\to\pi^0\nu\bar\nu)$ has been
proposed recently by Buchalla and Buras \cite{bb}.

\item[ii)] If one determines $\acpmi(\bkk)$ from measurements of
the corresponding CP-violating asymmetries and $\bar\vartheta$, $\vartheta$
from the two-triangle constructions outlined in refs.~\cite{hlgr}-\cite{lon},
the hadronic uncertainties arising in (\ref{e15}) and (\ref{e17}) from
QCD-penguins with internal $u$- and $c$-quarks affecting the
extraction of the CKM-angle $\beta$ can be eliminated and
another clean determination of $\beta$ is possible (up
to corrections related to $SU(3)$-breaking and certain neglected
diagrams).

\end{itemize}

In summary, we have shown that the CP-violating asymmetries
$\acpdir(\bkk)$ and $\acpmi(\bkk)$ are generated within the framework
of the Standard Model by QCD-penguins with
internal up- and charm-quark exchanges and are, thus, interesting
quantities to obtain experimental insights into the physics of these
contributions. Estimates obtained by applying the
perturbative approach of Bander, Silverman and Soni give rather
promising asymmetries of the order $(10-50)\%$ depending strongly
on the angle $\beta$ and branching ratios at the $10^{-6}$ level.
Therefore, measured non-vanishing CP asymmetries would not necessarily
imply physics beyond the Standard Model as claimed in
several previous papers.

While the direct CP-violating asymmetry $\acpdir(\bkk)$ should be
equal to the one arising in the charged $B$-decay $B^+\to K^+\bar
K^0$, the mixing-induced CP asymmetry $\acpmi(\bkk)$ can be predicted
by using triangle relations among $B$-meson decay-amplitudes which
follow from the $SU(3)$ flavour symmetry of strong interactions and
certain dynamical assumptions. These assumptions consist, e.g., of
neglecting annihilation-like topologies. Uncertainties related to
$SU(3)$-breaking effects have been discussed in \cite{grl}-\cite{bf}.
Additional corrections to the triangle relations could also arise from
electroweak penguins which may, in the presence of a heavy top-quark,
lead to sizable contributions to the penguin sectors of $B$-decays
into final states containing mesons with CP-self-conjugate quark
contents \cite{dun}-\cite{dh}. Therefore, experimental tests of the
validity of the triangle relations (\ref{e13}) and (\ref{e14}) are
desirable.

In the future, when it will hopefully be possible to measure
CP-violating asymmetries arising in the decay $\bkk$, the results presented
in this letter should provide such a test of the $SU(3)$ triangle
relations and, moreover, of our understanding of CP violation.

\vspace{0.6cm}

It is a great pleasure to thank Andrzej Buras for constant
encouragement, useful discussions and a careful reading of the
manuscript.

\newpage

\section*{Figure Captions}
\begin{table}[h]
\begin{tabular}{ll}
Fig.\ 1:&The dependence of $\acpmi(\bkk)$ on $k^2/m_b^2$
for $R_t=1$ and\\
&various values of the CKM-angle $\beta$.\\
&\\
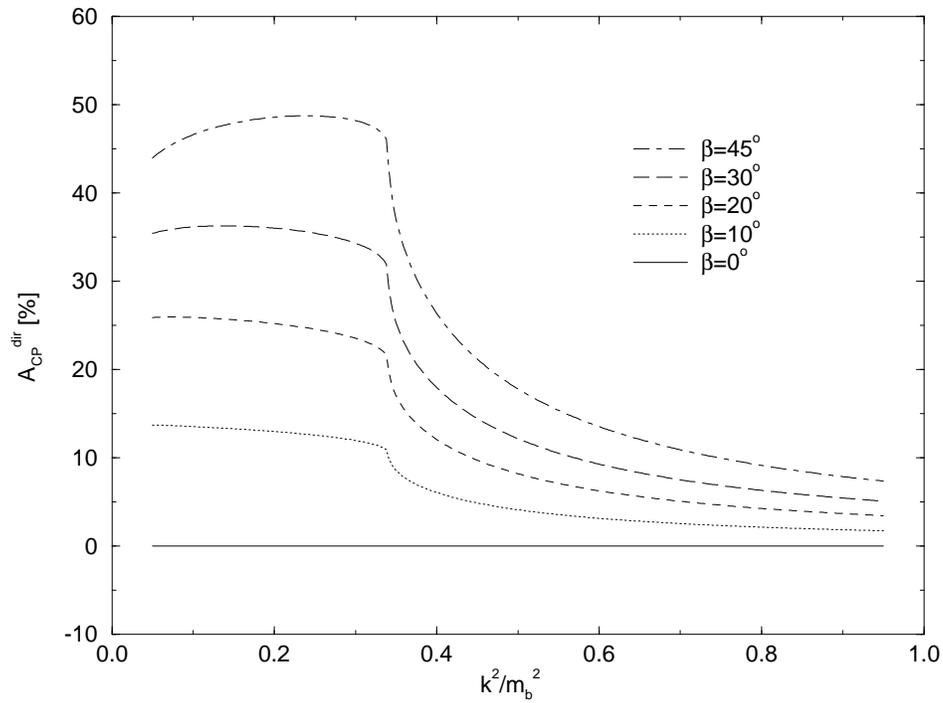
Fig.\ 2:&The dependence of $\acpdir(\bkk)$ on $k^2/m_b^2$
for $R_t=1$ and\\
&various values of the CKM-angle $\beta$.\\
&\\
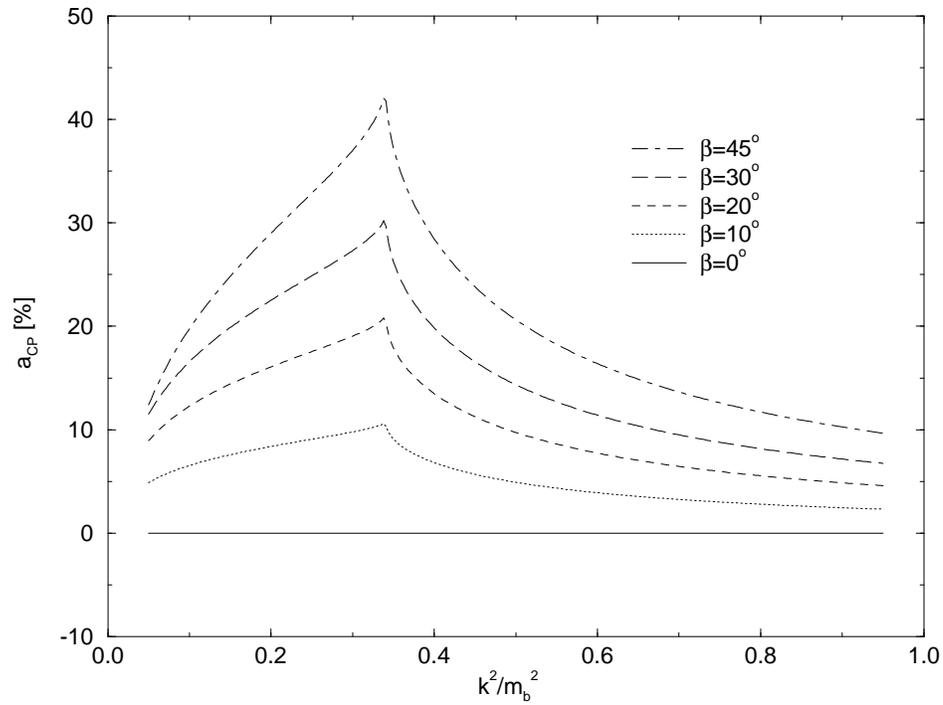
Fig.\ 3:&The dependence of the time-integrated CP asymmetry $\acp$ of the\\
&decay $\bkk$ on $k^2/m_b^2$ for $R_t=1$ and various values of the\\
&CKM-angle $\beta$. ($x_d=0.7$)\\
&\\
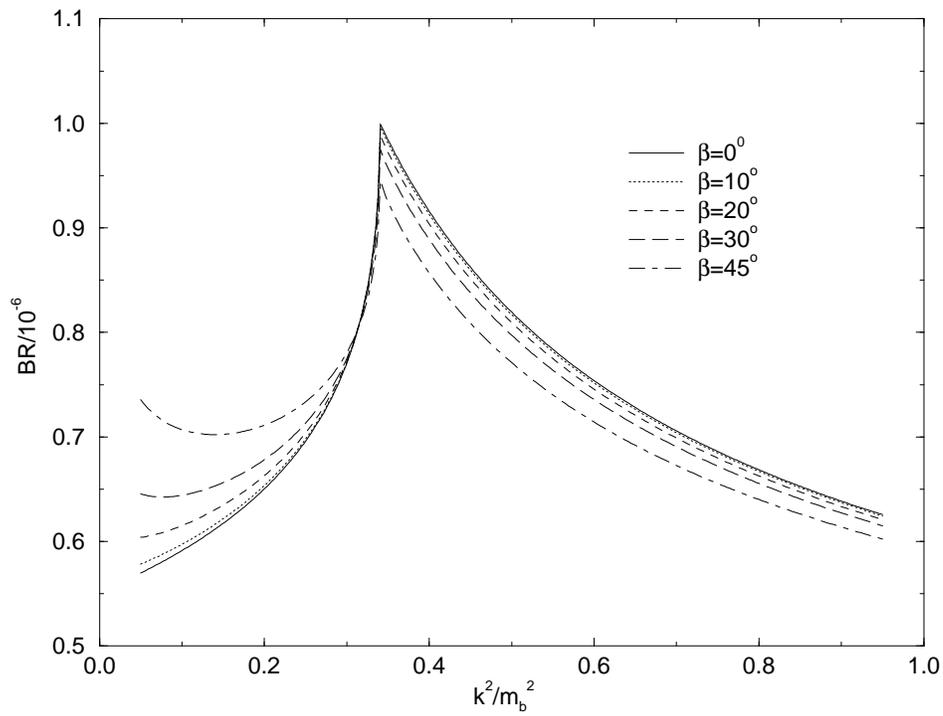
Fig.\ 4:&The dependence of the ``average'' branching ratio BR$(\bkk)$\\
&defined by eq.~(\ref{e12a}) on $k^2/m_b^2$ for $R_t=1$ and various
values of the\\
&CKM-angle $\beta$. ($\tau_{B_d}=1.6$ ps,
$|V_{cb}|=0.04$, $\Lambda^{(4)}_{\overline{\mbox{{\scriptsize
MS}}}}=0.3$ GeV)\\
\end{tabular}
\end{table}

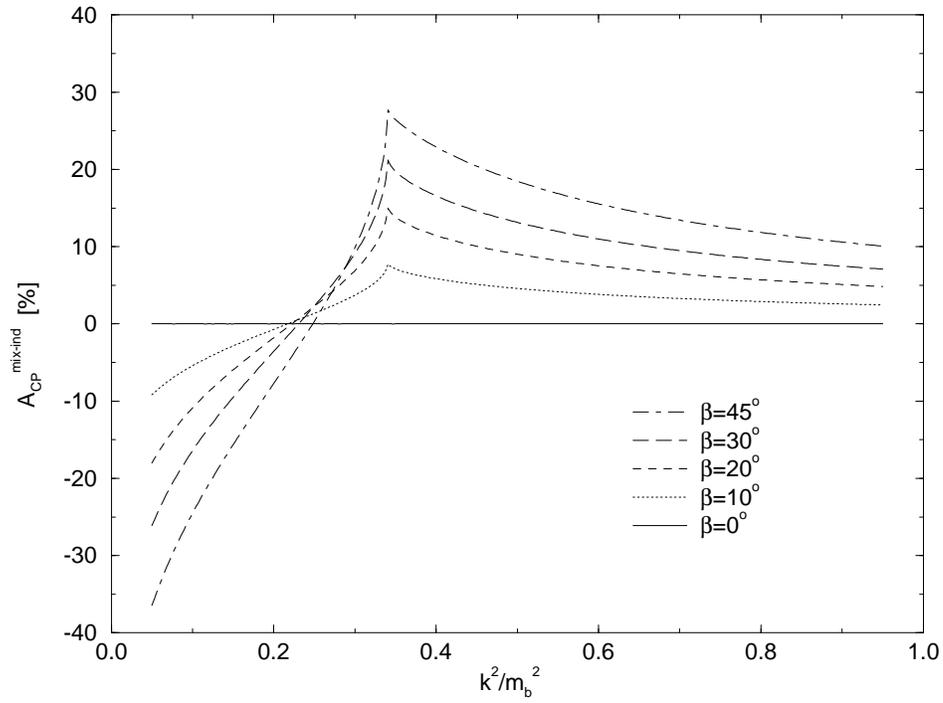
\end{document}